\documentclass[a4paper,fleqn]{cas-sc}

\usepackage[numbers]{natbib}

\usepackage{acro}
\usepackage{mathtools}
\usepackage{optidef}
\usepackage{siunitx}
\usepackage{booktabs}

\newdefinition{remark}{Remark}
\newdefinition{assumption}{Assumption}

\DeclareAcronym{bla}{
  short = BLA,
  long  = \emph{best linear approximation}
}
\DeclareAcronym{nrmse}{
  short = NRMSE,
  long  = normalized root mean square error
}
\DeclareAcronym{mse}{
  short = MSE,
  long  = mean square error
}
\DeclareAcronym{rmse}{
  short = RMSE,
  long  = root mean square error
}
\DeclareAcronym{nllfr}{
  short = NL-LFR,
  long  = \emph{nonlinear linear fractional representation}
}
\DeclareAcronym{pnlss}{
  short = PNLSS,
  long  = polynomial nonlinear state-space
}
\DeclareAcronym{rms}{
  short = RMS,
  long  = root mean square
}
\DeclareAcronym{lti}{
  short = LTI,
  long  = linear time-invariant
}

\DeclareAcronym{pispo}{
  short = PISPO,
  long  = \emph{periodic in, same period out}
}
\DeclareAcronym{ode}{
  short = ODE,
  long  = ordinary differential equation
}

\DeclareAcronym{zoh}{
  short = ZOH,
  long  = zero-order hold
}

\DeclareAcronym{snr}{
  short = SNR,
  long  = signal-to-noise ratio
}
\DeclareAcronym{dft}{
  short = DFT,
  long  = discrete Fourier transform
}
\DeclareAcronym{idft}{
  short = IDFT,
  long  = inverse discrete Fourier transform
}

\DeclareAcronym{frf}{
  short = FRF,
  long  = frequency response function
}

\DeclareAcronym{frm}{
  short = FRM,
  long  = frequency response matrix
}

\DeclareAcronym{rk4}{
  short = RK4,
  long  = fourth-order Runge-Kutta
}

\DeclareAcronym{sdof}{
  short = SDOF,
  long  = single degree of freedom
}

\DeclareAcronym{mdof}{
  short = MDOF,
  long  = multi degree of freedom
}

\DeclareAcronym{rfs}{
  short = RFS,
  long  = restoring force surface
}

\begin{document}
\let\WriteBookmarks\relax
\def\floatpagepagefraction{1}
\def\textpagefraction{.001}
\shorttitle{}
\shortauthors{Merijn Floren and Jan Swevers}

\title[mode = title]{A sliding-window approach for latent restoring force modeling}                      
\tnotemark[1]

\tnotetext[1]{This work is supported by Flanders Make's IRVA projects ASSISStaNT and CoMoDO.}

\author[1,2]{Merijn Floren}[orcid=0000-0001-7265-7851]
\cormark[1]

\credit{Conceptualization, Formal analysis, Methodology, Software, Validation, Visualization, Writing -- original draft preparation}

\author[1,2]{Jan Swevers}[orcid=0000-0003-2034-5519]

\credit{Funding acquisition, Resources, Supervision, Writing -- review and editing}

\affiliation[1]{organization={KU Leuven, Department of Mechanical Engineering},
                city={Leuven},
                postcode={3001}, 
                country={Belgium}}

\affiliation[2]{organization={Flanders Make@KU Leuven, MPRO Core Lab},
                city={Leuven},
                postcode={3001}, 
                country={Belgium}}

\cortext[cor1]{Corresponding author, e-mail: \texttt{merijn.floren@kuleuven.be}.}

\begin{abstract}
Restoring force surface (\acs{rfs}) methods offer an attractive nonparametric framework for identifying nonlinear restoring forces directly from data, but their reliance on complete kinematic measurements at each degree of freedom limits scalability to multidimensional systems.
The aim of this paper is to overcome these measurement limitations by proposing an identification framework with relaxed sensing requirements that exploits periodic multisine excitation.
Starting from an initial linear model, a sliding-window feedback approach reconstructs latent states and nonlinear restoring forces nonparametrically, enabling identification of the nonlinear component through linear-in-parameters regression instead of highly non-convex optimization.
Validation on synthetic and experimental datasets demonstrates high simulation accuracy and reliable recovery of physical parameters under partial sensing and noisy conditions.
\end{abstract}



\maketitle

\section{Introduction}
Identifying state-space models of mechanical systems from input-output data is inherently challenging, as the system dynamics depend on latent states that are not directly measurable. As a result, parameter estimation typically requires solving a highly nonlinear and non-convex optimization problem with an intrinsic recurrent structure imposed by the state evolution. However, this difficulty can be significantly alleviated if an estimate of the latent state trajectory can be obtained separately from the parameter estimation step. In that case, the recurrent nature of the identification problem effectively disappears, reducing parameter estimation to a considerably simpler regression problem.

The above procedure illustrates the fundamental rationale underlying \ac{rfs} methods.
As a concrete example, consider the \ac{sdof} system studied in the original \ac{rfs} formulation~\cite{masri1979nonparametric}:
\begin{equation}
m\ddot{y}(t) + f\big(y(t), \,\dot{y}(t)\big) = u(t),
\label{eq:rest_orig}
\end{equation}
where \(m\) denotes the mass, \(y(t)\) the displacement, 
\(u(t)\) the external excitation, and \(f(\cdot)\) the total restoring force, representing the combined linear and nonlinear stiffness and damping forces acting on the system. In the original work~\cite{masri1979nonparametric}, it is assumed that both the excitation \(u(t)\) and the acceleration \(\ddot{y}(t)\) are measured (or otherwise available), and that the mass \(m\) is known or can be reliably estimated. Under these assumptions, the restoring force can be isolated through simple algebraic rearrangement:
\begin{equation}
f\big(y(t),\,\dot{y}(t)\big) = u(t) - m\ddot{y}(t).
\label{eq:rest_isolated}
\end{equation}
The right-hand side of~\eqref{eq:rest_isolated} is therefore directly computable, enabling a fully nonparametric reconstruction of the restoring force behavior. The corresponding arguments of \(f(\cdot)\) are obtained from the measured acceleration \(\ddot{y}(t)\) through numerical integration, effectively reconstructing the latent states needed to avoid recursive parameter estimation.
With these quantities available, the restoring force can be approximated using any suitable nonlinear function approximator\footnote{The original study~\cite{masri1979nonparametric} employed polynomial basis functions, while later work also considered neural-network representations~\cite{masri1993identification}.}. Crucially, this modeling step is performed in a static regression setting, where the restoring force is treated as an explicit algebraic mapping from \(y(t)\) and \(\dot{y}(t)\) to the computed force values, i.e., the right-hand side of~\eqref{eq:rest_isolated}. As a result, identifying \(f(\cdot)\) reduces to a supervised function approximation problem with independent regression samples, rather than a recursive and non-convex dynamical estimation problem.

Although first introduced in 1979~\cite{masri1979nonparametric}, \ac{rfs} methods remain relevant today; the survey in~\cite{noel2017tenyears} recognizes them as a major time-domain identification technique, valued for their simplicity and intuitive, visual interpretations. \ac{rfs} methods have, for example, been used to characterize the complex nonlinear forces at the wing-to-payload mounting interface in an F-16 fighter jet~\cite{dossogne2015nonlinear}. Other applications include predicting the response of elastomer materials~\cite{saad2006equivalent}, analyzing the variable stiffness of an elastomagnetic suspension~\cite{bonisoli2007identification}, and identifying nonlinear behavior in a robotic arm \cite{goge2006experiences}.

\subsection{Limitations of classical \ac{rfs} methods}
Despite its success, the original \ac{rfs} method exhibits several limitations. First, it relies on knowledge of acceleration, velocity, and displacement. Ideally, all three quantities should be measured directly; however, this is often impractical or prohibitively expensive. In most applications, only one quantity is measured, while the remaining variables are obtained through numerical processing, a procedure that can introduce significant errors~\cite{worden1990data}. In mechanical systems, acceleration is typically measured, with velocity and displacement obtained via numerical integration. This procedure tends to amplify low-frequency noise and sensor drift over time, often requiring additional post-processing. Conversely, if displacement is measured, velocity and acceleration must be obtained through numerical differentiation, which rapidly becomes unreliable in the presence of measurement noise. A detailed discussion of the implications of numerical integration and differentiation for \ac{rfs}-based approaches can be found in~\cite{worden1990data}.

A second limitation of \ac{rfs} methods arises from their demanding measurement requirements in higher-dimensional systems. Specifically, extending the \ac{sdof} restoring-force isolation in~\eqref{eq:rest_isolated} to the \ac{mdof} case requires measurements at every degree of freedom~\cite{masri1982non, masri1982nonparametric}. In practice, for high-dimensional systems, instrumenting all degrees of freedom quickly becomes impractical, as sensors may alter system dynamics, compromise structural integrity, be physically inaccessible, or result in excessive costs.

Although numerical post-processing and increased sensor coverage can partially alleviate these limitations, they do not fundamentally remove the strict measurement dependence of \ac{rfs} methods. Recent work has therefore sought to address this issue more directly. In~\cite{rogers2022latent}, a Bayesian formulation is proposed in which a linear system of \acp{ode} is driven by a Gaussian process in time representing the unknown nonlinear component of the restoring force, enabling joint inference of both the latent state trajectory and the nonlinear restoring force. A related deterministic method was introduced in~\cite{floren2022nonlinear}, where the nonlinear restoring force time series is analytically reconstructed from an initially fitted linear model using a sliding-window procedure. Both approaches, however, explicitly assume an \ac{sdof} system formulation and therefore do not extend directly to the \ac{mdof} setting.

\subsection{Proposed approach}
This work extends the rationale introduced in~\cite{floren2022nonlinear} to the \ac{mdof} setting, thereby substantially relaxing the measurement requirements of \ac{rfs} methods in higher-dimensional systems.
In particular, no restrictions are imposed on the type of measured
quantity (displacement, velocity, or acceleration), nor on the number
or spatial locations of the measured degrees of freedom\footnote{Provided that
the measured responses are dynamically coupled to the applied excitation.}.
Following~\cite{floren2022nonlinear}, the only requirement is that the system is excited using random-phase multisine signals, which enable estimation of the \ac{bla} from
frequency response measurements~\cite{pintelon2012system}.
This initial linear model is subsequently employed within a sliding-window framework to analytically reconstruct both the nonlinear restoring force and the associated latent state trajectory in the time domain. While the spatial location of the nonlinear restoring force is assumed to be known, its functional form and magnitude remain unknown. The linear dynamics are assumed to follow a known \ac{ode} structure, whereas the physical parameters of this linear subsystem are only approximately known through available initial estimates. The objective is twofold: to recover the true physical parameters of the underlying linear system and to obtain an accurate nonlinear model suitable for simulation and control applications.

\subsection{Notation}
The sets of real, integer, and natural numbers are denoted by \(\mathbb{R}\), \(\mathbb{Z}\), and \(\mathbb{N}\), respectively.
The identity matrix is denoted by \( I \), and \(0\) denotes the zero matrix, with dimensions clear from context.
For a real-valued vector \(x \in \mathbb{R}^n\) and a symmetric positive definite matrix \(Q \in \mathbb{R}^{n \times n}\), the squared weighted 2-norm is defined as \(\|x\|_Q^2 = x^\mathrm{T} Q x\), where \(x^\mathrm{T}\) denotes the transpose of \(x\). When a unit weighting is used, the notation simplifies to \(\|x\|^2 = x^\mathrm{T} x\). For complex-valued vectors, the same definition applies except that the Hermitian transpose, denoted by \(x^\mathrm{H}\), replaces the regular transpose. The imaginary unit is denoted by \(j\). 
The exponential of a square matrix \(A\) is written as \(\exp(A)\).

\subsection{Paper outline}
The remainder of this paper is structured as follows. Section~\ref{sec:ch3_problem} defines the problem statement, after which Section~\ref{sec:ch3_method} details the proposed sequential identification procedure. The methodology is validated using three case studies: a simulated \ac{sdof} system in Section~\ref{subsec:duffing_case_study}, a simulated \ac{mdof} system with unmeasured nonlinear locations in Section~\ref{sec:results_mdof}, and experimental validation on an \ac{sdof} benchmark setup in Section~\ref{sec:results_silverbox}. Conclusions are drawn in Section~\ref{sec:ch3_conclusion}.

\section{Problem statement}\label{sec:ch3_problem}
\begin{figure}
    \centering
    \includegraphics[scale=1.25]{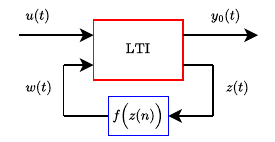}
    \vspace{6pt}
    \caption{The \ac{nllfr} model structure as a feedback interconnection of a \ac{lti} system and a static nonlinear mapping.
    }
    \label{fig:ch3_modstruct}
\end{figure}

We consider the \ac{pispo} system class, comprising systems for which a periodic input signal yields a steady-state output with the same period\footnote{The \ac{pispo} class encompasses a broad range of systems, including those exhibiting amplitude-dependent resonance and nonlinearities such as saturation and dead zones, while excluding phenomena that produce subharmonics, such as bifurcations and chaotic behavior.}.
The governing \acp{ode} are assumed to follow a known structural form, with the exception of an unknown nonlinear restoring force, and are represented within the \ac{nllfr} framework, in which nonlinear systems are described as a feedback interconnection between an \ac{lti} dynamical system and a static nonlinear mapping, as illustrated in Fig.~\ref{fig:ch3_modstruct}. The corresponding continuous-time system dynamics are described by the following state-space equations:
\begin{subequations}
    \begin{align}
      \dot{x}(t) &= \mathcal{A} x(t) + \mathcal{B}_u u(t) + \mathcal{B}_w w(t), \label{eq:xt2}\\
      y_0(t) &= \mathcal{C}_y x(t) + \mathcal{D}_{yu} u(t) + \mathcal{D}_{yw} w(t), \label{eq:yt2}\\
      z(t) &= \mathcal{C}_z x(t), \label{eq:zt2}\\
      w(t) &= f\big(z(t)\big), \label{eq:wt2}
    \end{align}
    \label{eq:plant2}%
\end{subequations}
where \(x(t) \in \mathbb{R}^{n_x}\) is the latent state, \(u(t) \in \mathbb{R}^{n_u}\) the external input, \(w(t) \in \mathbb{R}^{n_w}\) the nonlinear feedback input\footnote{That is, the nonlinear restoring force.}, \(y_0(t) \in \mathbb{R}^{n_y}\) the exact output, and \(z(t) \in \mathbb{R}^{n_z}\) the latent argument of the nonlinear mapping \(f: \mathbb{R}^{n_z} \to \mathbb{R}^{n_w}\). 
The matrices \(\mathcal{A} \in \mathbb{R}^{n_x \times n_x}\), \(\mathcal{B}_u \in \mathbb{R}^{n_x \times n_u}\), \(\mathcal{B}_w \in \mathbb{R}^{n_x \times n_w}\), \(\mathcal{C}_y \in \mathbb{R}^{n_y \times n_x}\), \(\mathcal{C}_z \in \mathbb{R}^{n_z \times n_x}\), \(\mathcal{D}_{yu} \in \mathbb{R}^{n_y \times n_u}\), and \(\mathcal{D}_{yw} \in \mathbb{R}^{n_y \times n_w}\) describe the linear dynamics and input-output relationships. Since the linear system of \acp{ode} is known, the structure of these matrices is also known. However, the values of the corresponding physical parameters \(\theta_{\text{phys}}\) remain unknown, although an initial guess is available, denoted as \(\theta_{\text{phys},0}\).
 The true parameters \(\theta_{\text{phys}}\) correspond to the \emph{underlying linear system}, which {exclusively} captures all linear input-output dynamics. Accordingly, the nonlinear mapping \(f(\cdot)\) in~\eqref{eq:wt2} is purely nonlinear\footnote{
For instance, \(f(z)\) cannot be expressed as \(f(z) = K z + g(z)\), where \(K\) is a constant matrix and \(g(\cdot)\) is a nonlinear function.
}.

Data are collected from~\eqref{eq:plant2} by exciting the system with random-phase multisine inputs~\cite{pintelon2012system}:
\begin{equation}
    u(n) \coloneqq \frac{2}{\sqrt{N}} \sum_{k=1}^{N/2 - 1} U_k \cos(2\pi k f_0 n + \varphi_k),
    \label{eq:multisine}
\end{equation}
where \(U_k\) denotes the excitation amplitude at frequency index \(k\), and \(f_0 = f_s/N\) is the frequency resolution, with \(f_s\) denoting the sampling frequency and \(N\) the (even) number of samples per period. The corresponding frequency of the \(k\)th harmonic is given by \(f_k = k f_0\). The phases \(\varphi_k\) are independent and uniformly distributed over the interval \([0, 2\pi)\).

\begin{assumption}\label{ass:poe}
The input signal \(u(n)\) is persistently exciting of sufficiently high order, such that all relevant system dynamics are excited over the frequency band of interest.
\end{assumption}
\begin{assumption}\label{ass:zoh}
The discrete-time signal~\eqref{eq:multisine} is applied to the continuous-time system~\eqref{eq:plant2} via \ac{zoh}, i.e., \( u_{\mathrm{ZOH}}(t) \coloneqq u(n)\), for \( t \in [nT_s, (n+1)T_s) \), where \(T_s = 1/f_s\) denotes the sampling period.
\end{assumption}
\begin{assumption}\label{ass:noise}
The measured output is corrupted by additive, zero-mean, stationary noise \(v(n)\) with finite variance, i.e., \(y(n) \coloneqq y_0(n) + v(n)\). The noise may be colored and is uncorrelated with the input \(u(n)\), which is exactly known.
\end{assumption}

The resulting input-output dataset for parameter estimation consists of \(P \geq 1\) steady-state periods of \(R \geq n_u\) realizations of a random-phase multisine~\eqref{eq:multisine}:
\begin{equation}
    \mathcal{D} = \big\{\big(u^{\left[r,p\right]}(n),\, y^{\left[r,p\right]}(n)\big)\big\}_{n=0,r=0,p=0}^{N-1,R-1,P-1},
    \label{eq:train_data}
\end{equation}
where both the input and output signals are zero-mean.
Based on \(\mathcal{D}\), the objective is to identify a discrete-time \ac{nllfr} state-space model of the form:
\begin{subequations}
    \begin{align}
      {x}(n+1) &= {A} x(n) + {B}_u u(n) + {B}_w w(n),\label{eq:mod_x}\\
      y(n) &= {C}_y x(n) + {D}_{yu} u(n) + {D}_{yw} w(n),\label{eq:mod_y}\\
      z(n) &= {C}_z x(n),\label{eq:mod_z}\\
      w(n) &= \beta^\mathrm{T} \phi\big(z(n)\big),\label{eq:mod_w}
    \end{align}
    \label{eq:model2}%
\end{subequations}
where the discrete-time matrices in~\eqref{eq:mod_x} are related to their continuous-time counterparts in~\eqref{eq:xt2} through the \ac{zoh} discretizations:
\begin{equation}
    A = \exp(\mathcal{A} T_s), \qquad
    B_u = \int_0^{T_s} \exp(\mathcal{A} \tau) \mathcal{B}_u \, d\tau, \qquad
    B_w = \int_0^{T_s} \exp(\mathcal{A} \tau) \mathcal{B}_w \, d\tau,
    \label{eq:zoh_discretisation}
\end{equation}
while \(C_y=\mathcal{C}_y\), \(C_z=\mathcal{C}_z\), \(D_{yu}=\mathcal{D}_{yu}\), and \(D_{yw}=\mathcal{D}_{yw}\).
Moreover, \(\phi: \mathbb{R}^{n_z} \to \mathbb{R}^{n_{\phi}}\) represents a polynomial feature mapping with \(\beta \in \mathbb{R}^{n_{\phi} \times n_w}\) the corresponding coefficient matrix. Optimization is thus carried out over the decision variables \(\theta = ({\theta}_{\text{phys}},\, {\beta})\), which preserves the physical interpretability of the underlying linear system, despite the final model being expressed in discrete time. An implication of this choice is that the discrete-time matrices have to be recomputed at each iteration of the optimization routine that updates \(\theta_{\text{phys}}\). Fortunately, this can be done efficiently using a single matrix exponential~\cite{van2003computing}:
\begin{equation}
    \begin{bmatrix}
        A & B \\
        0 & I
    \end{bmatrix} =
    \exp\left(
        \begin{bmatrix}
            \mathcal{A} & \mathcal{B} \\
            0 & 0
        \end{bmatrix}
        T_s
    \right),\label{eq:smart_zoh}
\end{equation}
where \(\mathcal{B} = [\mathcal{B}_u\; \mathcal{B}_w]\) and \(B = [B_u \; B_w]\).
In the remainder of this paper, all discrete-time matrices are obtained explicitly via this formulation.

\begin{remark}\label{rem:zoh_discretisation} The \ac{zoh} discretization method in \eqref{eq:zoh_discretisation} assumes that both inputs \(u(t)\) and \(w(t)\) are piecewise constant over each sampling interval \(t \in [nT_s,\, (n+1)T_s)\) for all \(n\). Under ideal experimental conditions, this assumption holds for the external excitation \(u(t)\), but not for the internal feedback signal \(w(t)\), which varies continuously. Consequently, the sampled continuous-time \ac{nllfr} \emph{system} in~\eqref{eq:plant2} cannot be represented exactly by the discrete-time \emph{model} in~\eqref{eq:model2}. Nevertheless, by choosing a sufficiently small sampling period \(T_s\), the induced approximation error can be kept within acceptable bounds.
\end{remark}

\section{A step-wise identification algorithm}\label{sec:ch3_method}
A three-step identification algorithm is proposed. First, in Section~\ref{subsec:init_linear}, the initial physical parameters \(\theta_{\text{phys}}\) are optimized to accurately capture the linearized input-output behavior. Second, in Section~\ref{subsec:rest_force}, the nonlinear restoring force is inferred nonparametrically, after which the parameters \(\beta\) are obtained by solving a linear system of equations, while keeping \(\theta_{\text{phys}}\) fixed. Third, in Section~\ref{subsec:final_opti}, all parameters \(\theta\) are jointly optimized to compensate for bias introduced in the preceding steps and to further improve the model's simulation accuracy.

\subsection{Initial linear model}\label{subsec:init_linear}
\begin{figure}
  \centering
  \includegraphics[scale=1.25]{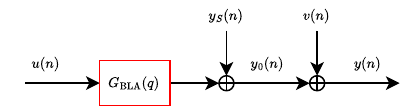}
  \vspace{6pt}
  \caption{Schematic overview of the \ac{bla} framework. For random excitation signals with a Gaussian distribution, the response of a nonlinear system is replaced by the sum of an \ac{lti} approximation \(G_{\text{BLA}}(q)\), unmodeled nonlinear dynamics \(y_{S}(n)\), and a disturbing noise source \(v(n)\).}
  \label{fig:bla_schematic}
\end{figure}
The \ac{pispo} system class and the assumed experimental conditions in Section~\ref{sec:ch3_problem} allow us to use the theory of the \ac{bla}~\cite{pintelon2012system, enqvist2005linear}. When applied with random excitation signals that follow a Gaussian distribution, such as the random-phase multisine~\eqref{eq:multisine}, this framework represents the response of a nonlinear dynamical system as the sum of an \ac{lti} approximation, unmodeled nonlinear dynamics, and a disturbing noise source. A schematic representation of this concept is provided in Fig.~\ref{fig:bla_schematic}. The \ac{bla} itself describes a linearized dynamic relationship between the zero-mean input \(u(n)\) and the zero-mean output \(y(n)\) that is optimal in the mean-square sense:
\begin{equation}
  {G}_{\text{BLA}}(q) \coloneqq \arg \min_{G} \mathbb{E}\left[ \, \|y(n) - G(q)u(n)\|^2 \,\right],
  \label{eq:bla_definition}
\end{equation}
where \(\mathbb{E}\left[\cdot\right]\) denotes the expectation operator and \(q\) the forward-shift operator. As shown in \cite{pintelon2012system}, for the single-input case and under multisine excitation, the minimizer of~\eqref{eq:bla_definition} is equivalent to
\begin{equation}
  {G}_{\text{BLA}}(j\omega_k) =  \mathbb{E}\left[\,  \frac{Y(k)}{U(k)} \,\right],
  \label{eq:bla_frequency_domain}
\end{equation}
where \(Y(k)\) and \(U(k)\) denote the leakage-free \acp{dft} of \(y(n)\) and \(u(n)\), respectively, at frequency line \(k\), therefore \(\omega_k = 2\pi kf_0\). The expectation is taken over different periods and realizations in \(\mathcal{D}\).

In our setting, the nonparametric \ac{bla} provides an averaged \ac{frm} that facilitates the subsequent parametrization of the model components directly related to the \ac{nllfr} input-output behavior, i.e., the state-space matrices \(\mathcal{A}\), \(\mathcal{B}_u\), \(\mathcal{C}_y\), and \(\mathcal{D}_{yu}\).
Furthermore, the \ac{bla} estimation procedure yields an assessment of the nonlinear distortions and the disturbing noise level, offering valuable insight into the system's behavior. These quantities can also be incorporated into the parameter estimation procedure as weighting functions. For more details on the \ac{bla} estimation procedure and its properties, we refer to~\cite{pintelon2012system, enqvist2005linear}.

Let \(\hat{G}_{\text{BLA}}(j\omega_k)\) denote the nonparametric estimate of the \ac{bla} at frequency \(\omega_k\). The corresponding parametric \ac{bla}, parametrized by \(\theta_{\text{phys}}\), is given by
\begin{equation}
    G(\zeta_k \mid \theta_{\text{phys}}) = C_y \big(\zeta_k I - A\big)^{-1} B_u + D_{yu},
    \label{eq:par_bla}
\end{equation}
where \(\zeta_k\) denotes the z-transform variable evaluated on the unit circle at frequency \(f_k = k f_0\).
Starting from the initial guess \(\theta_{\text{phys},0}\), the physical parameters \(\theta_{\text{phys}}\) are optimized by minimizing the weighted squared residuals between the nonparametric and parametric \ac{bla} estimates over the excited frequency lines:
\begin{mini}[1]
    {\theta_{\text{phys}}}
    {\frac{1}{|\mathcal{K}|} \sum_{k \in \mathcal{K}}\big\|W(k) \odot \big(\hat{G}_{\text{BLA}}(j\omega_k) -  {G}(\zeta_k \mid \theta_{\text{phys}})\big) \big\|_F^2,\label{eq:bla_par_loss}}
    {}{}
\end{mini}
where \(\|\cdot\|_F^2\) represents the squared Frobenius norm, \(\odot\) the element-wise Hadamard product, and \(\mathcal{K} \subseteq \{1, \dots, N/2 - 1\}\) the set of excited frequency lines, with \(|\mathcal{K}|\) denoting its cardinality.
Moreover, \(W(k)\) denotes a frequency-dependent weighting matrix, which, in principle, can be any well-chosen real-valued matrix. Yet, we adopt the common choice of weighting the squared model residuals by the reciprocal of the total variance of the nonparametric \ac{bla} estimates (see~\cite[Ch.~4.3.1]{pintelon2012system}). Such weighting effectively implements a whitening transformation that accounts for unequal variance for unequal variance across frequency lines. If the total variance estimate is not available, i.e., in case of a single multisine experiment, a unity weighting is applied instead.

\begin{remark}
  In the parametrization of the \ac{bla} in~\eqref{eq:bla_par_loss}, each excited frequency line is treated independently within the optimization problem. This decoupling eliminates the recursion-induced non-convexity typically encountered in time-domain parameter estimation, leading to a substantially more tractable optimization problem.
\end{remark}
\begin{remark}
  The nonparametric \ac{bla} is affected by a systematic error due to nonlinear distortions~\cite{pintelon2012system}. Consequently, the nonparametric \ac{bla} is biased, and optimizing \(\theta_{\text{phys}}\) by minimizing the model residual in~\eqref{eq:bla_par_loss} results in biased parameter estimates. This bias conflicts with the objective of identifying the true physical parameters. Nevertheless, at this stage, the primary objective is to achieve accurate simulation performance. The resulting bias will be addressed and compensated for in the final step of the proposed algorithm, described in Section~\ref{subsec:final_opti}.
\end{remark}

\subsection{Modeling the nonlinear restoring force}\label{subsec:rest_force}
In this second step, the nonlinear restoring force and the latent state trajectories are identified for fixed \(\theta_{\text{phys}}\). To improve both the \ac{snr} and the computational efficiency we proceed with the sample means \(u^{[r]}(n) = \frac{1}{P} \sum_{p=0}^{P-1} u^{[r,p]}(n)\) and \(y^{[r]}(n) = \frac{1}{P} \sum_{p=0}^{P-1} y^{[r,p]}(n)\).
Note that under the \ac{zoh} assumption, averaging over the inputs is redundant as they are noise-free, i.e., \(u^{[r]}(n) = u^{[r,p]}(n)\). Nevertheless, we include this step here to accommodate the case where measured (and thus noisy) input data are used instead.

\subsubsection{A nonparametric sliding window feedback approach}\label{subsec:nonpar_restoring}
When simulating the output of the parametric \ac{bla} model and comparing it to the measured output, a mismatch arises due to the unmodeled nonlinear dynamics. The idea of this step is to obtain a nonparametric estimate of the nonlinear feedback force \(w\) that improves the agreement between the simulated and measured outputs. 
To this end, we propose a sliding window strategy, in which the nonlinear feedback force is estimated by solving a local optimization problem over a finite time horizon that is shifted forward at each sample instant.
Specifically, the nonlinear feedback force is estimated over a window of length \(H +1 \ll N\), with \(H \in \mathbb{N}\) being the {prediction horizon} length. Only the solution corresponding to the current time instance is retained and used to shift the window forward in time by one sample. This procedure is repeated until all \(N\) time steps have been processed. Along the way, the full latent state \(x\) is naturally inferred as well.

We first define the stacked vectors:
\begin{equation}
      \mathcal{U}^{[r]}(n) = \begin{bmatrix}
          u^{[r]}(n) \\
          u^{[r]}(n+1) \\
          \vdots \\
          u^{[r]}(n+H)
      \end{bmatrix}, \qquad
      \mathcal{Y}^{[r]}(n) = \begin{bmatrix}
          y^{[r]}(n) \\
          y^{[r]}(n+1) \\
          \vdots \\
          y^{[r]}(n+H)
      \end{bmatrix}, \qquad
      \mathcal{W}^{[r]}(n) = \begin{bmatrix}
          w^{[r]}(n) \\
          w^{[r]}(n+1) \\
          \vdots \\
          w^{[r]}(n+H)
      \end{bmatrix},
    \label{eq:stacks}%
\end{equation}
with dimensions \(\mathbb{R}^{(H+1)n_u}\), \(\mathbb{R}^{(H+1)n_y}\), and \(\mathbb{R}^{(H+1)n_w}\), respectively.
Then, the optimization problem that is solved for each time step \(n \in \{0,\, \dots,\, N-1\}\) and each realization \(r \in \{0,\, \dots,\, R-1\}\) is given by
\begin{mini!}
    {\mathcal{W}^{[r]}(n)}{
        \frac{1}{2}\big\| \mathcal{Y}^{[r]}(n) -\mathcal{\hat{Y}}^{[r]}(n) \big\|^2_{\mathcal{Q}} 
        + \frac{\lambda}{2} \big\| \mathcal{W}^{[r]}(n) \big\|^2,\label{eq:mhe_cost}}
    {\label{eq:mhe_time}}{}
    \addConstraint{\mathcal{\hat{Y}}^{[r]}(n)}{=\mathcal{O}_x x^{[r]}_*(n) + \mathcal{S}_u \mathcal{U}^{[r]}(n) + \mathcal{S}_w \mathcal{W}^{[r]}(n),}{\label{eq:stacked_sim_outs}}
\end{mini!}
where \(\lambda \in \mathbb{R}_{>0}\) is a regularization parameter that controls the solution variability, and \(\mathcal{Q}\) is a block-diagonal matrix formed by concatenating \(H+1\) copies of the inverse of the time-domain sample noise covariance matrix defined in~\eqref{eq:sample_noise_cov_time}\footnote{
In cases where the sample noise covariance matrix is not available, 
\(\mathcal{Q}\) is constructed as a diagonal matrix containing the inverses of the output variances.
Doing so ensures that, in the multi-output case, the different outputs are properly scaled relative to their magnitudes.}. Equation~\eqref{eq:stacked_sim_outs} expresses the stacked simulated output as a function of the initial state \(x^{[r]}_*(n)\), which will be defined shortly,
and the stacked inputs \(\mathcal{U}^{[r]}(n)\) and \(\mathcal{W}^{[r]}(n)\); the evolution of this sequence is described with the following block-matrices:
\begin{equation}
      \mathcal{O}_x = \begin{bmatrix}
          C_y \\
          C_y A \\
          \vdots \\
          C_y A^H
      \end{bmatrix},\;
      \mathcal{S}_u=\begin{bmatrix}
          D_{yu} & 0 & \cdots & 0 \\
          C_y B_u & D_{yu} & \cdots & 0\\
          \vdots & \vdots & \ddots & \vdots\\
          C_y A^{H-1}B_u & C_y A^{H-2}B_u & \cdots & D_{yu} 
      \end{bmatrix},\;
      \mathcal{S}_w = \begin{bmatrix}
          D_{yw} & 0 & \cdots & 0 \\
          C_y B_w & D_{yw} & \cdots & 0\\
          \vdots & \vdots & \ddots & \vdots\\
          C_y A^{H-1}B_w & C_y A^{H-2}B_w & \cdots & D_{yw}
      \end{bmatrix},
    \label{eq:block_matrices}%
\end{equation}
with dimensions \(\mathbb{R}^{n_y (H+1) \times n_x}\), \(\mathbb{R}^{n_y (H+1)\times n_u(H+1)}\), and \(\mathbb{R}^{n_y(H+1) \times n_w(H+1)}\), respectively. 

The optimization problem in~\eqref{eq:mhe_time} is convex. Therefore, by substituting~\eqref{eq:stacked_sim_outs} into~\eqref{eq:mhe_cost}, setting the gradient with respect to \(\mathcal{W}^{[r]}(n)\) to zero, and solving for \(\mathcal{W}^{[r]}(n)\), we obtain the closed-form solution:
\begin{equation}
    \mathcal{W}^{[r]}_*(n) = -\mathcal{G}^{-1} \mathcal{S}_w^\mathrm{T} \mathcal{Q} \big(\mathcal{O}_x x_*^{[r]}(n) + \mathcal{S}_u \mathcal{U}^{[r]}(n) - \mathcal{Y}^{[r]}(n)\big),
    \label{eq:opti_W}
\end{equation}
where \( \mathcal{G} = \mathcal{S}_w^\mathrm{T} \mathcal{Q} \mathcal{S}_w + \lambda I\).

\begin{remark}\label{rem:regularisation}
The regularization term in~\eqref{eq:mhe_cost} is introduced to address two practical issues. First, the measured output data \(\mathcal{Y}^{[r]}(n)\) are corrupted by noise, and without regularization (i.e., when \(\lambda = 0\)) the minimizer of~\eqref{eq:mhe_time}
would be highly sensitive to this noise, resulting in large variability in the inferred \(\mathcal{W}^{[r]}_*(n)\).
Second, when \(n_w > n_y\), the unregularized optimization problem is inherently ill-posed (i.e., \(\mathcal{G}\) is rank-deficient), leading to non-unique solutions; the regularization term in~\eqref{eq:mhe_cost} ensures that a unique minimizer exists in such cases.
The hyperparameter \(\lambda\) controls the trade-off between data fit and solution variability.
\end{remark}

\begin{remark}\label{rem:window_length}
    The main motivation for using a prediction horizon length greater than one, or a longer window in general, lies in its ability to mitigate the effects of measurement noise. When the window is short, \(\mathcal{W}^{[r]}_*(n)\) may become overly sensitive to high-frequency noise in \(\mathcal{Y}^{[r]}(n)\), thus resulting in a nervous estimate. In contrast, longer windows naturally lead to smoother solutions, as the individual noisy samples are averaged out over time.
    The effect of the prediction horizon length \(H\) and regularization parameter \(\lambda\), including their sensitivity to different noise levels, is empirically studied in Section~\ref{sec:results_duffing}.
\end{remark}

Equation~\eqref{eq:opti_W} provides the optimal solution over the defined window, and is naturally defined for all \(R\) realizations provided that \(n < N-H\). That is, when \(n \geq N-H\), computing \(\mathcal{U}^{[r]}(n)\) and  \(\mathcal{Y}^{[r]}(n)\) requires ``future'' input-output data beyond the final sample instant.
However, thanks to periodicity, we are able to define the input-output data at every possible time step \(m\in \mathbb{Z}\) in terms of the first \(N\) samples as follows:
\begin{equation}
    u^{[r]}(m) \coloneqq u^{[r]}(m \bmod N), \qquad y^{[r]}(m) \coloneqq y^{[r]}(m \bmod N),
    \label{eq:modulus}
\end{equation}
where \(\bmod\) denotes the modulus operator.
Using~\eqref{eq:modulus}, the optimal solution in~\eqref{eq:opti_W} is well-defined at each time step and for every realization.
However, in~\eqref{eq:opti_W}, the initial state of each window, denoted \(x_*^{[r]}(n)\), is required. The recursive evolution of this state gives rise to the ``sliding'' nature of the proposed approach, as detailed next.

To advance to the next sample, we only use the first \(n_w\) elements of \(\mathcal{W}^{[r]}_*(n)\), which we denote with \({w}^{[r]}_*(n)\). These elements correspond to the current time step, and are related to the state according to
\begin{equation}
    {x}^{[r]}_*(n) = \begin{cases}
        x^{[r]}_0, & \text{if }n = 0 \\
        A x^{[r]}_*(n  -  1) + B_u u^{[r]}(n  - 1) + B_w w^{[r]}_*(n  -  1),\qquad& \text{otherwise}
    \end{cases},
    \label{eq:state_evo}
\end{equation}
which shows that a value for \(x^{[r]}_0\) is still needed at \(n = 0\). This initial state is typically unknown, and arbitrarily setting it (e.g., to zero) introduces a transient response that is incompatible with the true system, and, therefore, unsuitable for parametric modeling of the nonlinear restoring force. While simply discarding the corresponding transient samples is possible, it also results in the loss of valuable information.

Instead, the idea is to use the parametric \ac{bla} model to obtain a more informed initial state. In addition, the periodicity definition in~\eqref{eq:modulus} can be leveraged to retain all \(N\) samples. Specifically, the \ac{bla} state trajectories are computed in the frequency domain as
\begin{equation}
    X_{\text{BLA}}^{[r]}(k) = \big(\zeta_k I - A\big)^{-1} B_u U^{[r]}(k),
    \label{eq:bla_state_freq}
\end{equation}
where \(U^{[r]}(k)\) denotes the \ac{dft} of \(u^{[r]}(n)\) at frequency line \(k\).
Taking the inverse \ac{dft} (\acs{idft}) of \(X_{\text{BLA}}^{[r]}(k)\) yields \(x_{\text{BLA}}^{[r]}(m) \coloneqq x_{\text{BLA}}^{[r]}(m \bmod N)\) for \(m \in \mathbb{Z}\). This periodicity is essential, as it allows the simulation to be initialized at an arbitrary phase of the same periodic trajectory. Concretely, in~\eqref{eq:state_evo} we define \(x_0^{[r]} \coloneqq x_{\text{BLA}}^{[r]}(-N_0)\) and \(u^{[r]}(0) \coloneqq u^{[r]}(-N_0)\), where \(N_0 \in \mathbb{N}_0\) denotes the number of offset samples required for the transient to decay. The simulation is then run for a total of \(N + N_0\) samples, after which the first \(N_0\) samples of the resulting sequences are discarded.

In summary, the proposed procedure described in this section yields an analytical expression for the missing nonlinear restoring force, which can be evaluated efficiently over the entire time series. For given values of the horizon length \(H\), regularization strength \(\lambda\), and offset length \(N_0\), advancing the simulation one sample at a time yields a set of paired samples as
\begin{equation}
    \mathcal{D}_{wz} = \big\{\big(w^{\left[r\right]}_*(n),\, z^{\left[r\right]}_*(n)\big)\big\}_{n=0,r=0}^{N-1,R-1},
    \label{eq:train_data_wz}
\end{equation}
where \(z^{\left[r\right]}_*(n)\) is easily derived from \(x^{\left[r\right]}_*(n)\) through~\eqref{eq:mod_z}. This dataset will be used in the next step for parametric modeling of the nonlinear restoring force.

\subsubsection{Parametric modeling of the nonlinear restoring force}
Given \(\mathcal{D}_{wz}\), estimating the nonlinear coefficient matrix \(\beta\) reduces to solving a set of linear regression problems. Due to the black-box nature of the polynomial basis function model, the individual entries of \(\beta\) do not necessarily carry a direct physical interpretation. Nevertheless, when \(n_w > 1\), it is important that \(\beta\) respects the physical location of the nonlinearities.

To enforce this structure, we introduce a binary selection matrix \(P_{z,i} \in \{0,1\}^{n_{z,i} \times n_z}\), which, for each location \(i \in \{1, \ldots, n_w\}\), extracts the correct subvector of dimension \(n_{z,i} \leq n_z\) from \(z\). The corresponding nonlinear restoring force term is then modeled as
\begin{equation}
    w_i(n) = \beta_i^\mathrm{T} \phi_i\big(P_{z,i} z(n)\big),
    \label{eq:w_i}
\end{equation}
where \(\phi_i : \mathbb{R}^{n_{z,i}} \to \mathbb{R}^{n_{\phi,i}}\) is a monomial feature map, and \(\beta_i \in \mathbb{R}^{n_{\phi,i}}\) contains the associated coefficients. 

The full nonlinear mapping in~\eqref{eq:mod_w} is subsequently defined as a decoupled combination of the individual nonlinearities through
\begin{equation}
    \beta \coloneqq \begin{bmatrix}
        \beta_1 & \ldots & 0 \\
        \vdots  & \ddots & \vdots \\
        0 & \ldots & \beta_{n_w}
    \end{bmatrix} \in \mathbb{R}^{n_{\phi} \times n_w}, \quad
    \phi \coloneqq \begin{bmatrix}
        \phi_1 \\
        \vdots \\
        \phi_{n_w}
    \end{bmatrix} : \mathbb{R}^{n_z} \to \mathbb{R}^{n_{\phi}},
\end{equation}
with \(n_\phi = \sum_{i=1}^{n_w} n_{\phi,i}\). Since~\eqref{eq:w_i} is linear in the unknown parameters, each vector \(\beta_i\) can be estimated analytically using ordinary least squares on the input-target pairs from \(\mathcal{D}_{wz}\), aggregated over all \(R\) realizations.
Doing so yields the estimated nonlinear coefficient matrix \(\beta\), which, together with \(\theta_{\text{phys}}\), fully initializes the \ac{nllfr} model.

\begin{remark}\label{rem:lin_monomials}
  The bias in the \(\theta_{\text{phys}}\) estimate has an important impact on the structure of the monomials in \(\phi_i(\cdot)\). Specifically, the restoring force estimates \(w_*\) inferred through~\eqref{eq:mhe_time} absorb the discrepancy between the true linear state trajectories and those reconstructed using the estimated linear parameters. As a result, the inferred force contains an additional component that depends approximately linearly on the reconstructed state trajectories. Consequently, \(\phi_i(\cdot)\) must include degree-one monomials to accurately represent the mapping from \(z_*\) to \(w_*\), even though the true restoring force in~\eqref{eq:wt2} corresponds to a purely nonlinear mapping in \(z\). This discrepancy motivates the bias-correction procedure and enables recovery of the true linear parameters in the subsequent step.
\end{remark}

\subsection{Final optimization}\label{subsec:final_opti}
The previous steps provided fully initialized estimates of the parameters \(\theta = (\theta_{\text{phys}},\, \beta)\).
In this final step, the goal is twofold: (i) to enhance the simulation performance and (ii) to recover the true system parameters.  These objectives are addressed by solving the following optimization problem:
\begin{mini}[1]
    {\theta}
    { \frac{1}{RN} \sum_{r=0}^{R-1} \sum_{k=0}^{N/2} \| Y^{[r]}(k) - \hat{Y}^{[r]}(k \mid \theta)\|^2_{W(k)}
                + \gamma \| {\beta}^{[1]}(\theta)\|_1.\label{eq:dual_cost}}
    {}{}
\end{mini}
The first component in~\eqref{eq:dual_cost} quantifies the simulation error in the frequency domain. 
Here, \(Y^{[r]}(k)\) is the measured output spectrum, computed by applying the \ac{dft} to \(y^{[r]}(n)\), while  \(\hat{Y}^{[r]}(k \mid  \theta)\) is the modeled output spectrum, obtained by simulating the \ac{nllfr} model in~\eqref{eq:model2} with input \(u^{[r]}(n)\) and then applying the \ac{dft} to the resulting time-domain output.
To avoid spectral leakage caused by transients from an unknown initial state, we implement the same procedure described in Section~\ref{subsec:nonpar_restoring}, i.e., simulating and subsequently discarding \(N_0\) offset samples to ensure steady-state conditions. The proposed frequency-domain evaluation allows the weighting matrix \(W(k)\) to account for the varying noise characteristics across frequencies. When available, the inverse of the sample noise covariance matrix in~\eqref{eq:sample_noise_cov_freq} is used at each frequency line. If this matrix is not available, \(W(k)\) is instead constructed as a diagonal matrix containing the inverses of the output variances at each frequency, ensuring that the samples are properly balanced both across frequencies and, in the multi-output case, across different outputs.

The second component in~\eqref{eq:dual_cost} serves as a regularization term aimed at recovering the true underlying linear parameters. As discussed in Remark~\ref{rem:lin_monomials}, the initial estimates of both \(\theta_{\text{phys}}\) and \(\beta\) include components associated with linear dynamics, which renders the former non-unique and therefore not physically interpretable. To address this ambiguity, the elements of \({\beta}\) that correspond to first-degree monomials are stacked into vector \({\beta}^{[1]}(\theta)\) and penalized using the sparsity-promoting \(\ell_1\)-norm, denoted by \(\|\cdot\|_1\). This strategy effectively redirects the contribution of \({\beta}^{[1]}(\theta)\) into \({\theta}_{\text{phys}}\), thereby encouraging the recovery of the true physical values. The strength of the proposed regularization term is controlled through hyperparameter \(\gamma \in \mathbb{R}_{\geq 0}\).

\subsection{Implementation details}
All algorithms are implemented in Python. The optimization routines are formulated using \texttt{JAX}~\cite{jax2018github} together with \texttt{Equinox}~\cite{kidger2021equinox}, enabling automatic differentiation and efficient execution on both CPUs and GPUs. Nonlinear least-squares problems are solved using the Levenberg-Marquardt algorithm~\cite{levenberg1944method,marquardt1963algorithm} as provided by the \texttt{Optimistix} library~\cite{optimistix2024}.

\section{Simulation studies}\label{sec:results_duffing}
Two simulation examples are presented to demonstrate the proposed step-wise algorithm. The first considers an \ac{sdof} Duffing oscillator and analyzes the influence of the different training steps, hyperparameters, and noise levels on model performance. The second examines a more complex \ac{mdof} system under limited sensing conditions.

\subsection{SDOF Duffing oscillator}\label{subsec:duffing_case_study}
In this section we study in detail the proposed step-wise algorithm on a simulation example of a forced Duffing oscillator, of which the dynamics are described by the following \ac{ode}:
\begin{equation}
    m \ddot{y}_0(t) + c \dot{y}_0(t) + k y_0(t) + k_3 y_0^3(t) = u(t),
    \label{eq:duffing_oscillator}
\end{equation}
with \(m\), \(c\) and \(k\) the linear mass, damping and stiffness parameters, respectively, and \(k_3\) the cubic stiffness parameter. Here, the exact noise-free output \(y_0(t)\) describes the displacement of the mass, while the input \(u(t)\) acts as an external forcing term applied to the system. If we define \(x(t)=\left[y_0(t),\, \dot{y}_0(t)\right]^\mathrm{T}\), the Duffing dynamics in~\eqref{eq:duffing_oscillator} adhere to the \ac{nllfr} structure of~\eqref{eq:plant2}, with
\begin{equation}
    \begin{aligned}
        \mathcal{A} &= \begin{bmatrix}
            0 & 1 \\
            -{k}/{m} & -{c}/{m}
        \end{bmatrix}, 
        &\quad \mathcal{B}_u &= -\mathcal{B}_w = \begin{bmatrix}
            0 \\
            {1}/{m}
        \end{bmatrix},\\
        \mathcal{C}_y &= \mathcal{C}_z = \begin{bmatrix}
            1 & 0
        \end{bmatrix}, 
        &\quad \mathcal{D}_{yu} &= \mathcal{D}_{yw} = 0,
    \end{aligned}
    \label{eq:second_order_ss}
\end{equation}
and the typically unknown \(f\big(z(t)\big)=k_3 z^3(t)\). Within this formulation, the physical parameter vector is defined as \(\theta_{\text{phys}}=(m,c,k)\).

\begin{table}
    \centering
    \caption{Parameter estimates of the Duffing oscillator obtained using the \ac{bla} procedure for 100 different initializations in~\eqref{eq:duf_init}. The estimates (mean \(\pm\) standard deviation) are highly consistent across initializations and noise levels, but bias is present, most noticeable in the stiffness parameter \(k\), which is systematically overestimated due to the hardening effect of the cubic spring.}
    {\fontsize{8.5}{8.5}\selectfont
    \begin{tabular}{l|ccc}
          & \(m\) [\SI{}{\kilogram}] & \(c\) [\SI{}{\newton\second\per\meter}] & \(k\) [\SI{}{\newton\per\meter}] \\
    \midrule
    true       & 1.00                                  & 2.00                         & 100 \\
    \ac{snr} \SI{60}{\decibel}  & \(0.988\) \(\pm\) 5.77\(\times10^{-6}\) & 2.10 \(\pm\) 6.34\(\times10^{-6}\)  & 114 \(\pm\) 6.44\(\times10^{-4}\) \\
    \ac{snr} \SI{40}{\decibel}  & 0.988 \(\pm\) 1.01\(\times10^{-5}\) & 2.10 \(\pm\) 8.07\(\times10^{-6}\)  & 114 \(\pm\) 1.13\(\times10^{-3}\) \\
    \ac{snr} \SI{20}{\decibel}  & 0.987 \(\pm\) 1.53\(\times10^{-6}\) & 2.10 \(\pm\) 5.77\(\times10^{-6}\)  & 114 \(\pm\) 1.76\(\times10^{-4}\)  
    \end{tabular}
    }
    \label{tab:duffing_parameters}
\end{table}

We generate synthetic input-output data by solving the Duffing equation~\eqref{eq:duffing_oscillator} using the \ac{rk4} integration scheme. The parameters of the Duffing oscillator are set to \(m = \SI{1}{\kilogram}\), \(c = \SI{2}{\newton\second\per\meter}\), \(k = \SI{100}{\newton\per\meter}\), and \(k_3 = \SI{500}{\newton\per\meter\cubed}\).
The training data consists of \(R = 5\) realizations of \(P = 2\) steady-state periods of a random-phase multisine~\eqref{eq:multisine}, with \(N = 8192\) samples each. The multisine input signal excites frequencies up to \SI{10}{\hertz}, is sampled at \(f_s = \SI{128}{\hertz}\), and has individual amplitudes \(U_k\) chosen such that an overall \ac{rms} amplitude of \SI{12}{\newton} is obtained. 
Starting from noise-free samples, we generate three datasets with increasing output noise levels, ranging from almost noise-free to strongly corrupted. Specifically, white Gaussian noise is added to the noise-free data to achieve \acp{snr} of \SIlist{60;40;20}{\decibel}. From these \acp{snr}, lower bounds on the achievable relative simulation errors can be derived, corresponding to \SIlist{0.1;1;10}{\percent}, respectively\footnote{
These bounds follow directly from the definition \(\text{SNR} = 10 \log_{10}(P_\mathrm{s}/P_\mathrm{n})\), where \(P_\mathrm{s}\) and \(P_\mathrm{n}\) denote the signal and noise power, respectively. The lower bound is then defined as the noise-to-signal amplitude ratio, i.e.\ \(\sqrt{P_\mathrm{n}}/\sqrt{P_\mathrm{s}} = 10^{-\text{SNR}/20}\), expressed as a percentage.
}.

The goal of this simulation example is threefold: first, to assess how each training step (\ac{bla} estimation, restoring force modeling, and final optimization) affects the parameter estimates and contributes to the model's simulation performance; second, to study the influence of the regularization strength \(\lambda\) and prediction horizon length \(H\) on the restoring force approach; and third, to evaluate the sensitivity of the above analyses with respect to the different levels of measurement noise.

\subsubsection*{Step \MakeUppercase{\romannumeral 1}: Best linear approximation}
\begin{table}
    \centering
    \caption{Simulation output \acp{nrmse} for the Duffing oscillator at different stages of the step-wise algorithm; compared to their respective theoretical lower bounds.
    The initial \ac{nllfr} models already show a significant improvement over the \ac{bla}; for \acp{snr} of \SIlist{60;40}{\decibel}, the lower bounds are not met, likely because the \ac{zoh} discretization error exceeds the respective noise levels.}
    {\fontsize{8.5}{8.5}\selectfont
    \begin{tabular}{lcccc}
    \toprule
    & \multicolumn{4}{c}{NRMSEs [\SI{}{\percent}]}  \\
    \cmidrule(lr){2-5}
     & \ac{bla} & initial \ac{nllfr} & optimized \ac{nllfr} & lower bound\\
    \midrule
    \ac{snr} \SI{60}{\decibel} & 20.5  & 4.38  & 1.08  & 0.10\\
    \ac{snr} \SI{40}{\decibel} & 20.5 & 4.51  & 1.48  & 1.00\\
    \ac{snr} \SI{20}{\decibel} & 22.8  & 10.8 & 10.0  & 10.0\\
    \bottomrule
    \end{tabular}
    }
    \label{tab:duffing_stepwise_performance}
\end{table}
The first step consists of estimating the parameters of the linear part of the model, as outlined in Section~\ref{subsec:init_linear}.
The concrete outcome is a linearized state-space model with estimates of the parameters \(m\), \(c\), and \(k\). To assess the sensitivity of the parameter estimates to initialization, we repeat the optimization procedure 100 times, each with a maximum of 100 Levenberg-Marquardt iterations, starting from different initial guesses generated as follows:
\begin{equation}
    \begin{bmatrix}
        m_0 \\
        c_0 \\
        k_0
    \end{bmatrix}
    =
    \begin{bmatrix}
        m(1 + \delta_m) \\
        c(1 + \delta_c) \\
        k(1 + \delta_k)
    \end{bmatrix},
    \qquad \text{where } \delta_m,\, \delta_c,\, \delta_k \sim \mathcal{U}(-0.9,\; 0.9), \label{eq:duf_init}
\end{equation}
with \(\mathcal{U}(-0.9,\; 0.9)\) denoting independent random variables sampled from a uniform distribution over the interval \([-0.9,\,0.9]\). In other words, each parameter is randomly initialized within \(\pm \SI{90}{\percent}\) of its true value.

The identification procedure was then carried out as described above, and the resulting parameter estimates are summarized in Table~\ref{tab:duffing_parameters}. Two main observations can be drawn from these results. First, the estimates are highly consistent across the 100 runs and noise levels, with small standard deviations indicating robust convergence to a unique solution. Second, the estimates are generally close to the true values, but a bias is present, as expected. This bias is most noticeable in the stiffness parameter \(k\), which is systematically overestimated due to the hardening effect of the cubic stiffness term \(k_3\). For further analyses, we proceed with the mean values of the estimates in Table~\ref{tab:duffing_parameters}.

\begin{figure}
    \centering
    \includegraphics[scale=1]{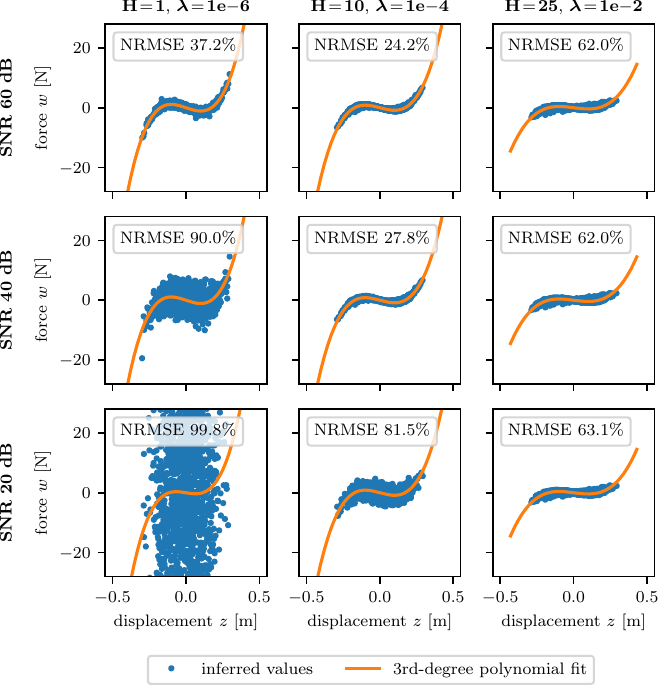
    }
    \caption{Visualization of \(\mathcal{D}_{wz}\) and corresponding polynomial fit for three hyperparameter combinations across all noise levels. The inferred values, obtained by solving~\eqref{eq:mhe_time}, generally suggest the correct cubic nature of the restoring force, but the influence of noise and hyperparameters is evident.}
    \label{fig:duffing_hyperparam_spring}
\end{figure}

We assess the simulation performance of the obtained \ac{bla} model using the normalized \ac{rms} error (\acs{nrmse}), computed here on the output signal\footnote{
    In general, the \ac{nrmse} is computed as the \ac{rms} value of the respective error signal divided by the \ac{rms} value of the reference signal, multiplied by \SI{100}{\percent}, thereby providing a relative and easily interpretable measure of performance.
}. The results, shown in Table~\ref{tab:duffing_stepwise_performance}, indicate a consistent \ac{nrmse} of around \SI{20}{\percent} across all noise levels. This relatively high error confirms that the linear model fails to adequately capture the Duffing oscillator's nonlinear behavior.

\subsubsection*{Step \MakeUppercase{\romannumeral 2}: Restoring force modeling}
In this second step, we address two main questions: (i) can the sliding-window approach suggest a restoring force \emph{signal} that drives the simulated output close to the desired output, and (ii) can a static nonlinear \emph{mapping} \(f\colon \mathbb{R}^{n_z} \to \mathbb{R}^{n_w}\) be obtained from this signal? Particular attention is paid to the influence of the regularization parameter \(\lambda\), the prediction horizon length \(H\), and the impact of measurement noise on the results. In the following, the offset length is set to \(N_0 = 100\) samples.
\begin{figure}
  \centering
  \includegraphics[scale=1]{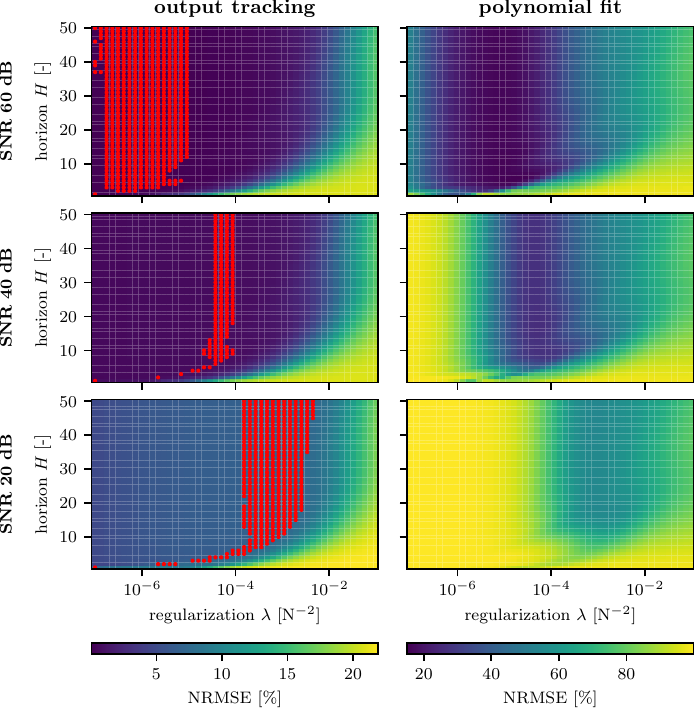}
    \caption{Grid search over the horizon length \(H\) and regularization strength \(\lambda\).
    The left column shows nonparametric \acp{nrmse} between measured and simulated output signals, while the right column shows parametric \acp{nrmse} of the corresponding polynomial fits. Red dots mark hyperparameter combinations where the nonparametric \acp{nrmse} are close to their lower bounds; these combinations also yield the most accurate polynomial fits.
    }
    \label{fig:duffing_grid_search}
\end{figure}

Figure~\ref{fig:duffing_hyperparam_spring} visualizes the inferred dataset \(\mathcal{D}_{wz}\) (in blue) for three hyperparameter settings across all noise levels, where the nonparametric restoring force \(w\) is plotted as a function of the displacement \(z\). Although not visible from the plots, in each case the inferred restoring force, obtained by solving~\eqref{eq:mhe_time}, has successfully driven the simulated output close to the desired output.
When plotted against displacement, this nonparametric force generally reveals the expected cubic nature of the underlying system nonlinearity.
However, the corresponding third-degree odd polynomial\footnote{Specifically, \(\phi(z)=[\;z,\; z^3]^{\mathrm{T}}\), as this structure matches the true underlying nonlinearity and includes the linear term required to compensate for the bias in the \({\theta}_{\text{phys}}\) estimate (see Remark~\ref{rem:lin_monomials}).
} fits (in orange) and their \acp{nrmse} indicate that the relationship between \(z\) and \(w\) does not always reflect a clear static mapping.
This phenomenon is most pronounced in the bottom-left plot, corresponding to the highest noise level combined with the shortest window length and weakest regularization. Here, the result lacks any meaningful structure, making it generally impossible to extract a reliable mapping from the signals in \(\mathcal{D}_{wz}\).

A more detailed analysis is carried out by performing a grid search over the hyperparameters. The prediction horizon length \(H\) is varied linearly from 1 to 50, while the regularization parameter \(\lambda\) is varied logarithmically from \(10^{-7}\) to \(10^{-1}\),
 using 50 values in total. The results are summarized in Fig.~\ref{fig:duffing_grid_search}, where the left column shows the nonparametric \acp{nrmse} between the measured and simulated outputs, while the right column presents the \acp{nrmse} of third-degree odd polynomial fits to the inferred restoring force signals, similar to Fig.~\ref{fig:duffing_hyperparam_spring}. The red dots mark the hyperparameter combinations for which the nonparametric output \acp{nrmse} are considered close\footnote{Defined using a relative tolerance of \SI{5}{\percent} and an absolute tolerance of \SI{0.02}{\percent}, with the larger of the two used as threshold.} to their lower bounds derived from the \ac{snr}. 
When examining \(\lambda\) first, Fig.~\ref{fig:duffing_grid_search} indicates that smaller values tend to reduce the nonparametric output \ac{nrmse}. However, this does not guarantee a good polynomial fit, since insufficient regularization causes the sliding-window algorithm to overfit the noise. The best polynomial fits are achieved at combinations that render the nonparametric output \acp{nrmse} close to their respective lower bounds. Here, regularization suppresses noise just enough without being too restrictive. 
As for the prediction horizon length \(H\), its effect is less pronounced, but it is clear that in general the window length should at least be a few samples long. Moreover, it can be observed that longer windows are preferred for higher noise levels. As discussed in Remark~\ref{rem:window_length}, this is because longer windows allow the algorithm to average out the high-frequency noise, leading to smoother estimates of the restoring force signal. Note that, in terms of computational cost, longer windows are somewhat more expensive to solve, but since the optimization problem admits an analytical solution, this increase is limited.

Finally, we assess the simulation performance of the initial \ac{nllfr} models obtained from the hyperparameter combination \(H=10\) and \(\lambda=10^{-4}\), i.e., the second column in Fig.~\ref{fig:duffing_hyperparam_spring}. The corresponding simulation output \acp{nrmse}, presented in Table~\ref{tab:duffing_stepwise_performance}, show a performance gain of more than a factor of four for the \acp{snr} of \SIlist{60;40}{\decibel}, but further improvement is still required. At the \ac{snr} of \SI{20}{\decibel}, the simulation accuracy is already close to the lower bound of \SI{10.0}{\percent}, despite the relatively large error observed in the polynomial fit of Fig.~\ref{fig:duffing_hyperparam_spring}. This polynomial model nonetheless captured the key cubic trend of the restoring force, which proves sufficient for accurate simulation. We proceed with the above initial \ac{nllfr} models to the next identification step.
\begin{figure}
    \centering
    \includegraphics[scale=1]{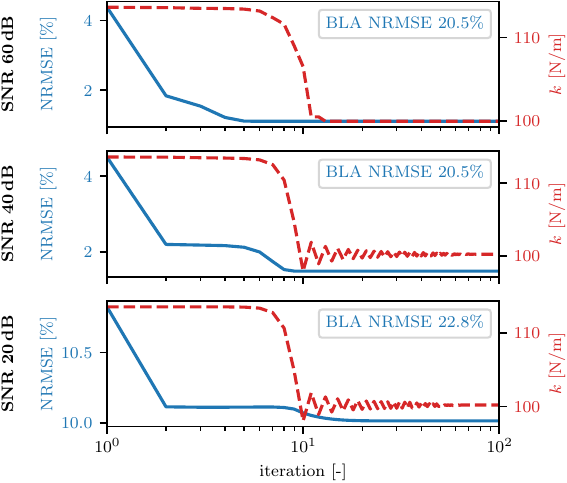
    }
    \caption{Evolution of the output simulation \ac{nrmse} and the linear stiffness parameter \(k\) over the iterations of the final optimization step. It can be seen that the optimization procedure first prioritizes reducing the simulation error, followed by correcting the bias in the stiffness parameter.}
    \label{fig:duffing_final_optimisation}
\end{figure}

\subsubsection*{Step \MakeUppercase{\romannumeral 3}: Final optimization}
The final identification step aims to improve the simulation performance of the initial \ac{nllfr} models, while simultaneously recovering the true parameters of the Duffing oscillator. This goal is achieved by minimizing the dual-objective cost function in~\eqref{eq:dual_cost} using 100 Levenberg-Marquardt iterations with a regularization parameter of \(\gamma = 5 \times 10^{-3}\). Figure~\ref{fig:duffing_final_optimisation} shows the evolution of the output \acp{nrmse} and the stiffness parameter \(k\) over the iterations. It can be observed that the optimization procedure initially focuses on reducing the \ac{nrmse}, and subsequently on correcting the parameter bias. For all three \acp{snr}, the \ac{nrmse} is significantly reduced, while the bias on \(k\) is almost entirely removed, as also apparent from the final parameter estimates in Table~\ref{tab:duffing_params_complete}.

The final output \acp{nrmse} in Table~\ref{tab:duffing_stepwise_performance} confirm a substantial improvement in simulation performance. However, for \acp{snr} of \SIlist{60;40}{\decibel}, the \acp{nrmse} remain above the ideal values expected from the noise levels. Since the model structure is correctly specified and noise is limited, this discrepancy is most plausibly due to the \ac{zoh} discretization, which does not exactly replicate the underlying continuous-time \ac{ode}, as discussed in Remark~\ref{rem:zoh_discretisation}. This discretization mismatch may also explain why not all parameters in Table~\ref{tab:duffing_params_complete} are recovered exactly.
\begin{table}
    \centering
    \caption{Physical parameter estimates of the Duffing oscillator after the final optimization step. 
    Compared with Table~\ref{tab:duffing_parameters}, the bias in the stiffness parameter is almost entirely eliminated.}
    {\fontsize{8.5}{8.5}\selectfont
    \begin{tabular}{l|cccc}
          & \(m\) [\SI{}{\kilogram}] & \(c\) [\SI{}{\newton\second\per\meter}] & \(k\) [\SI{}{\newton\per\meter}] & \(k_3\) [\SI{}{\newton\per\meter\cubed}] \\
    \midrule
    true       & 1.00           & 2.00  & 100 & 500 \\
    SNR 60 dB  & 0.999 & 2.06  & 99.9 & 485 \\
    SNR 40 dB  & 0.999 & 2.06  & 100 &  485\\
    SNR 20 dB  & 0.999 & 2.06  & 100 & 487
    \end{tabular}
    }
    \label{tab:duffing_params_complete}
\end{table}
\begin{figure}
    \centering
    \includegraphics[scale=1]{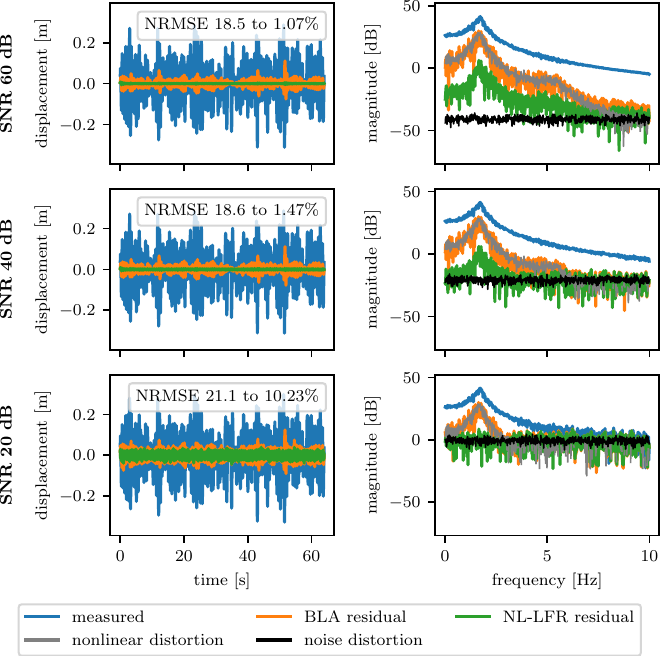}
    \caption{Test data: simulation performance of the Duffing oscillator \ac{bla} and \ac{nllfr} models in time and frequency domains across different noise levels. Although the \ac{nllfr} models outperform the \ac{bla}, the noise levels are not fully met for \acs{snr} of \SIlist{60;40}{\decibel}, most plausibly due to the \ac{zoh} discretization.}
    \label{fig:duffing_test}
\end{figure}

\subsubsection*{Performance on test data}
We conclude by evaluating the performance of the trained \ac{nllfr} models on unseen test data. The test set is a different random-phase realization of a multisine with the same properties as the training data. Figure~\ref{fig:duffing_test} presents the simulation results in both time and frequency domains for all noise levels.
The nonlinear and noise distortion levels are computed from the training data via the \ac{bla} estimation procedure described in~\cite[Ch.~4.3.1]{pintelon2012system}.
The performance is comparable to that on the training data, with time-domain output \acp{nrmse} similar to those in Table~\ref{tab:duffing_stepwise_performance}. In the frequency domain, optimal performance is, indeed, not achieved for \acp{snr} of \SIlist{60;40}{\decibel}, as the noise level is not met. 

The frequency-domain plots provide an additional validation of the \ac{bla} approach, as the residual between the measured response and the fitted linear model aligns with the nonparametrically estimated variance estimates. This observation indicates that the \ac{bla} captures the full linear contribution of the system dynamics, with the remaining mismatch attributable to nonlinear distortions and measurement noise.

\subsection{MDOF mass-spring-damper system}\label{sec:results_mdof}
In the previous simulation example, we considered an \ac{sdof} system where the nonlinearity depended directly on the measured output. The following simulation study investigates an \ac{mdof} system characterized by a nonlinear restoring force dependent on an unmeasured state. Under these conditions, classical \ac{rfs} approaches are not applicable.

\begin{figure}
    \centering
    \includegraphics[scale=1.25]{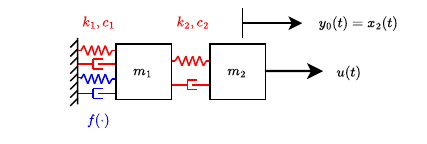}
    \caption{Schematic overview of the \ac{mdof} system. We excite and measure at the second mass, but the nonlinearity is between the first mass and the ground, which makes classical \ac{rfs} approaches inapplicable.}
    \label{fig:mdof_blockscheme}
\end{figure}

A schematic overview of the \ac{mdof} system is shown in Fig.~\ref{fig:mdof_blockscheme}.  
The system comprises two masses, \(m_1\) and \(m_2\), connected by linear springs and dampers, with the first mass linked to the ground via a nonlinear restoring force.  
The second mass serves as both the excitation and measurement point.
The corresponding physical parameter values are listed in Table~\ref{tab:mdof_params}.  
Defining \(z(t) = [x_1(t),\, \dot{x}_1(t)]^\mathrm{T}\), the nonlinear restoring force is modeled as the sum of a cubic spring and a damper with smooth saturation:  
\begin{equation}
    f\big(z(t)\big) = \alpha_1 \tanh\big(\alpha_2 z_2(t)\big) + \alpha_3 z_1^3(t),
    \label{eq:mdof_nonlinear_restoring_force}
\end{equation}
with parameters \(\alpha_1 = 7.0\), \(\alpha_2 = 3.0\), and \(\alpha_3 = 5.0 \times 10^4\).
The corresponding continuous-time state-space matrices in~\eqref{eq:plant2} can be obtained directly from the schematic in Fig.~\ref{fig:mdof_blockscheme} and are therefore omitted for brevity.

Synthetic, noiseless input-output data are generated by numerically solving the continuous-time system using the \ac{rk4} integration scheme. 
The training set comprises \(R = 6\) realizations, each containing \(P = 1\) steady-state period of a random-phase multisine~\eqref{eq:multisine}, with \(N = 8192\) samples per realization. 
The multisine input signal excites frequencies up to \SI{10}{\hertz}, is sampled at \(f_s = \SI{128}{\hertz}\), and its individual amplitudes \(U_k\) are chosen to yield an overall \ac{rms} amplitude of \SI{10}{\newton}. The output is defined as the displacement of the second mass.

\begin{figure}
    \centering
    \includegraphics[scale=1]{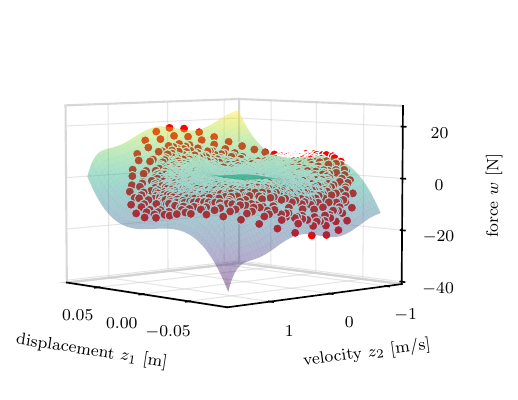}
    \caption{Nonlinear restoring force modeling for the \ac{mdof} mass-spring-damper system. The red dots represent the inferred nonparametric samples, while the smooth manifold represents the fitted polynomial model.}
    \label{fig:mdof_nonpar_rest}
\end{figure}

The initial values of the linear parameters are defined as
\begin{equation}
    \begin{bmatrix}
        m_{i,0} \\
        c_{i,0} \\
        k_{i,0}
    \end{bmatrix}
    =
    \begin{bmatrix}
        m_i(1 + \delta_{m_i}) \\
        c_i(1 + \delta_{c_i}) \\
        k_i(1 + \delta_{k_i})
    \end{bmatrix},
    \quad \text{where } \delta_{m_i},\, \delta_{c_i},\, \delta_{k_i} \sim \mathcal{U}(-0.9,\; 0.9),
\end{equation}
for \(i \in \{1,\,2\}\).
We perform 10 optimization runs with a maximum of 100 Levenberg-Marquardt iterations each, starting from different initial guesses, and proceed to the subsequent steps with the best performing model.
Next, for the modeling of the restoring force, we use \(H = 15\), \(N_0 = 100\), and \(\lambda = 10^{-8}\) in the nonparametric inference step.  
An odd polynomial of degree 7, without cross-terms, is then fitted to the inferred data.  
The results of both steps are shown in Fig.~\ref{fig:mdof_nonpar_rest}.

\begin{table}
    \centering
    \caption{True and estimated linear parameters of the \ac{mdof} mass-spring-damper system shown in Fig.~\ref{fig:mdof_blockscheme}. The value between parentheses indicates the effective linear damping coefficient that takes into account the contribution of the nonlinear damping term around zero velocity.
    }
    {\fontsize{8.5}{8.5}\selectfont
    \begin{tabular}{l|cccccc}
          & \(m_1\) [\SI{}{\kilogram}] & \(m_2\) [\SI{}{\kilogram}] & \(c_1\) [\SI{}{\newton\second\per\meter}] & \(c_2\) [\SI{}{\newton\second\per\meter}] & \(k_1\) [\SI{}{\newton\per\meter}] & \(k_2\) [\SI{}{\newton\per\meter}] \\
    \midrule
    true       & 2.00  & 1.00  & 5.00 (26.0) & 2.00 & 800 & 600 \\
    estimated  &  1.97 & 0.998  & 24.2 & 1.97 & 796  & 600
    \end{tabular}
    }
    \label{tab:mdof_params}
\end{table}

\begin{remark}
    During the nonparametric inference step, it is not necessary to specify whether the nonlinear restoring force originates from the spring, the damper, or a combination of both. Only the location of the nonlinearity needs to be specified. The precise nature of the nonlinearity can be decided later when fitting the polynomial model. This decision can be guided by visual inspection of a 3D plot such as Fig.~\ref{fig:mdof_nonpar_rest}, or through cross-validation over different candidate input configurations.
\end{remark}

\begin{figure}
    \centering
    \includegraphics[scale=1]{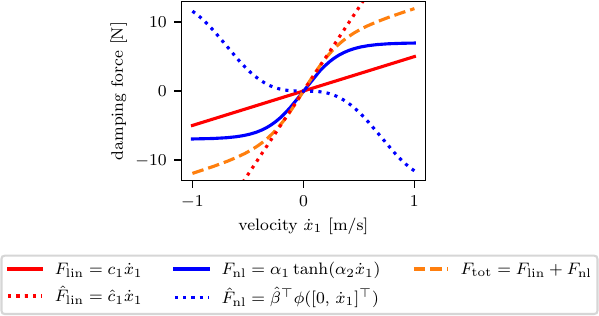}
    \caption{Total damping force decomposed into linear and nonlinear contributions.  
    The apparent overestimation of \(\hat{c}_1\) should be interpreted as the effective linear damping around zero velocity, which includes contributions from both the linear damper and the nonlinear saturation term.
    }
    \label{fig:damping_force_mdof}
\end{figure}

In the final optimization step, a maximum of 100 Levenberg-Marquardt iterations is performed with bias regularization parameter \(\gamma = 10^{-5}\).  
The resulting parameter estimates are listed in Table~\ref{tab:mdof_params}.  
Overall, the estimates are close to the true values, except for the damping coefficient \(c_1\), which appears to be overestimated by nearly a factor of five. This behavior is, however, expected given the explicit separation of the restoring force into linear and nonlinear components. In particular, the nonlinear saturation term
\(
F_{\mathrm{nl}} = \alpha_1 \tanh(\alpha_2 \dot{x}_1)
\)
in~\eqref{eq:mdof_nonlinear_restoring_force} exhibits a non-negligible linear contribution around zero velocity. 
Linearizing \(F_{\mathrm{nl}}\) with respect to \(\dot{x}_1\) at \(\dot{x}_1 = 0\) yields a slope of \(\SI{21.0}{\newton\second\per\meter}\), which, when added to the true linear damping \(c_1 = \SI{5.0}{\newton\second\per\meter}\), results in an effective linear damping of \(\SI{26.0}{\newton\second\per\meter}\). This value is significantly closer to the estimated damping coefficient of \(\SI{24.2}{\newton\second\per\meter}\).
This effect is further illustrated in Fig.~\ref{fig:damping_force_mdof}, which depicts the total damping force \(F_{\mathrm{tot}}\) as the sum of the linear contribution \(F_{\mathrm{lin}} = c_1 \dot{x}_1\) and the nonlinear saturation term \(F_{\mathrm{nl}} = \alpha_1 \tanh(\alpha_2 \dot{x}_1)\). The estimated effective linear damping \(\hat{c}_1\) aligns with the tangent of \(F_{\mathrm{tot}}\) in the low-velocity regime.

\begin{figure}
    \centering
    \includegraphics[scale=1]{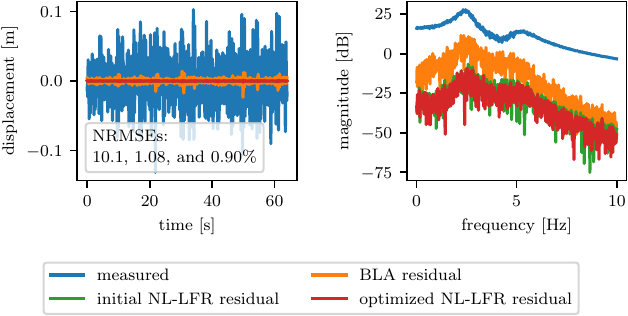}
    \caption{Test data: simulation performance of the \ac{mdof} mass-spring damper \ac{bla} and \ac{nllfr} models in time and frequency domains. The initial \ac{nllfr} model from the restoring force modeling step already performs close to the final model, which cannot be improved further due to modeling errors originating from the \ac{zoh} discretization.}
    \label{fig:mdof_test}
\end{figure}

We validate the trained \ac{nllfr} models on test data from a multisine realization with the same properties as the training set.  
Figure~\ref{fig:mdof_test} shows the simulation results in both the time and frequency domains.  
The final \ac{nllfr} model achieves an improvement of roughly a factor of ten compared to the \ac{bla} model.  
The figure also includes the initial \ac{nllfr} model from the restoring force modeling step, which already performs close to the final model, indicating that the final optimization mainly serves for bias correction. Further improvement is likely hindered by the \ac{zoh} discretization errors discussed previously.

\section{Experimental results on the Silverbox benchmark system}\label{sec:results_silverbox}
The proposed method is evaluated on the experimental data of the Silverbox benchmark~\cite{wigren2013three}, which represents an electronic implementation of the Duffing oscillator~\eqref{eq:duffing_oscillator}. In this experiment, the input voltage acts as an excitation analogous to mechanical force, whereas the measured output voltage represents the displacement response. The recorded data form an arrow-shaped trajectory consisting of two distinct parts and are sampled at approximately \SI{610}{\hertz}.
The first part, referred to as the ``arrowhead'' and used exclusively for testing, contains \num{40000} samples of a white Gaussian noise signal with gradually varying amplitude, filtered using a 9th-order Butterworth filter with cutoff frequency \SI{200}{\hertz}. The remaining data correspond to ten consecutive realizations of a random-phase multisine excitation, together comprising \num{86750} samples and exciting only odd harmonics up to \SI{200}{\hertz}. From this multisine portion, the final \num{21688} samples are reserved for testing purposes. Successive realizations are separated by short zero-input intervals and slightly exceed one full period length to accommodate transient samples, which are removed prior to analysis to ensure steady-state operation.
The data can be considered nearly noise-free~\cite{wigren2013three}; yet, it is not possible to compute the sample noise variance of the output, as each multisine realization consists of only a single period (see Appendix~\ref{sec:sample_noise_cov}).

The experimental Silverbox data deviate from ideal identification conditions in two respects. The input signal is measured after passing through a low-pass filter~\cite{wigren2013three}, such that it no longer satisfies the \ac{zoh} excitation of Assumption~\ref{ass:zoh}. 
Furthermore, the sampling frequency is only about three times higher than the maximum excited frequency. In system identification practice, sampling rates between ten and twenty times the highest frequency of interest are commonly recommended to accurately approximate continuous-time dynamics and to limit discretization errors. The relatively low sampling rate therefore introduces additional modeling inaccuracies. To alleviate these effects, the data are upsampled by a factor of twenty using cubic spline interpolation, after which simulation errors are evaluated on the downsampled signals.

\begin{figure}
    \centering
    \includegraphics[scale=1]{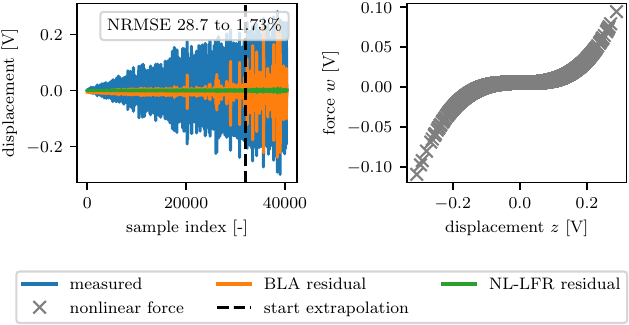}
    \caption{Simulation performance of the \ac{bla} and \ac{nllfr} models on the Silverbox arrowhead test data (left), and the corresponding nonlinear restoring force as simulated by the \ac{nllfr} model (right).}
    \label{fig:silverbox_test}
\end{figure}

As the Silverbox implements the dynamics of a Duffing oscillator, we assume the model structure in~\eqref{eq:second_order_ss}. We follow the step-wise procedure using the same optimization settings as in the previous sections. For \ac{bla} estimation, the initial values of the linear parameters are taken from~\cite{kocijan2018parameter}. In the sliding-window approach, we use a prediction horizon length of \( H = 10 \), an offset length \(N_0=100\), and a regularization strength of \( \lambda = 10^{-1} \). The static nonlinearity is modeled using a third-order polynomial, and a bias correction with \( \gamma = 10^{-1} \) is applied during the final optimization step. Both optimization steps used a maximum of 100 Levenberg-Marquardt iterations.

Figure~\ref{fig:silverbox_test} presents the simulation results on the arrowhead test data. The left plot shows the time-domain performance of the \ac{bla} and \ac{nllfr} models, while the right plot displays the corresponding nonlinear restoring force as simulated by the \ac{nllfr} model, which is clearly cubic. The \ac{nllfr} output simulation error is reduced by more than a factor of sixteen compared to the \ac{bla} model. The performance increase is also apparent from Table~\ref{tab:silverbox_results}, which shows the simulation errors of both models on both test datasets. It is worth noting that the \ac{nllfr} errors are very consistent, even in the extrapolation region, thanks to the correctly assumed nonlinear structure.

\begin{table}
    \centering
    \caption{\ac{bla} and \ac{nllfr} simulation errors on the Silverbox test datasets. 
    The \ac{nllfr} model remains accurate even in the extrapolation region of the arrowhead data thanks to its correctly assumed nonlinear structure.}
    {\fontsize{8.5}{8.5}\selectfont
    \begin{tabular}{lcccc}
    \toprule
    & \multicolumn{2}{c}{\ac{bla}} & \multicolumn{2}{c}{\ac{nllfr}} \\
    \cmidrule(lr){2-3} \cmidrule(lr){4-5}
     & NRMSE & RMSE & NRMSE & RMSE \\
    \midrule
    multisine & \SI{16.1}{\percent} & \SI{8.72}{\milli\volt} & \SI{1.54}{\percent} & \SI{0.825}{\milli\volt} \\
    arrowhead (full) & \SI{28.7}{\percent} & \SI{15.3}{\milli\volt} & \SI{1.74}{\percent}& \SI{0.927}{\milli\volt} \\
    arrowhead (no extrapolation) & \SI{19.7}{\percent} & \SI{8.47}{\milli\volt} & \SI{1.61}{\percent} & \SI{0.684}{\milli\volt}\\
    \bottomrule
    \end{tabular}
    }
    \label{tab:silverbox_results}
\end{table}
\begin{table}
    \centering
    \caption{\ac{bla} and \ac{nllfr} output simulation errors on the Silverbox test datasets when the data are upsampled by only a factor of five instead of twenty. Compared to Table~\ref{tab:silverbox_results}, the errors increase significantly, especially for the \ac{nllfr} model, highlighting the importance of proper upsampling.}
    {\fontsize{8.5}{8.5}\selectfont
    \begin{tabular}{lcccc}
    \toprule
    & \multicolumn{2}{c}{\ac{bla}} & \multicolumn{2}{c}{\ac{nllfr}} \\
    \cmidrule(lr){2-3} \cmidrule(lr){4-5}
     & NRMSE & RMSE & NRMSE & RMSE \\
    \midrule
    multisine & \SI{16.8}{\percent} & \SI{9.11}{\milli\volt} & \SI{5.25}{\percent} & \SI{2.85}{\milli\volt} \\
    arrowhead (full) & \SI{29.1}{\percent} & \SI{15.4}{\milli\volt} & \SI{5.89}{\percent}& \SI{3.14}{\milli\volt} \\
    arrowhead (no extrapolation) & \SI{20.3}{\percent} & \SI{8.70}{\milli\volt} & \SI{5.30}{\percent} & \SI{2.28}{\milli\volt}\\
    \bottomrule
    \end{tabular}
    }
    \label{tab:silverbox_results_low_samp}
\end{table}

The results in Table~\ref{tab:silverbox_results} outperform reported results in studies that identify physical model parameters from data, although there are not many physics-based methods to compare with. In~\cite{rogers2022latent}, a \SI{2.1}{\milli \volt} output \ac{rms} error (\acs{rmse}) was reported on the arrowhead data, while~\cite{kocijan2018parameter} reports a \SI{2.9}{\percent} output \ac{nrmse} on a fragment of the multisine data. 
As all three methods adopt the same model structure, the discrepancies between the proposed method and the results reported in~\cite{rogers2022latent} and~\cite{kocijan2018parameter} are likely attributable to differences in the discretization schemes used during optimization. This claim is supported by Table~\ref{tab:silverbox_results_low_samp}, which reports simulation errors on test data when the training data were upsampled by only a factor of five instead of twenty. For the \ac{nllfr} model in particular, the errors increased by more than a factor of three, emphasizing that proper upsampling is essential for simulation performance.

\section{Conclusion}\label{sec:ch3_conclusion}
This work introduced a novel approach to identifying and modeling the nonlinear restoring force in \ac{mdof} systems formulated as \ac{nllfr} state-space models.
 In contrast to traditional \ac{rfs} approaches, the proposed method relaxes the measurement assumptions considerably. In particular, any output quantity can be measured (displacement, velocity, or acceleration), only a subset of the degrees of freedom needs to be measured, and the measurements may be noisy.

Starting from an initial linear model, the nonlinear restoring force is reconstructed from measured data using a sliding-window approach. The resulting nonparametric estimate is subsequently used to identify the static nonlinear mapping within the \ac{nllfr} structure by fitting a polynomial basis function model. Since both the nonparametric inference and nonlinear parametrization admit closed-form solutions, the overall identification procedure remains computationally efficient and straightforward to implement. A final optimization stage improves simulation accuracy while compensating for any bias in the estimated physical parameters. Validation on simulated examples and experimental Silverbox data confirms accurate nonlinear model identification, with remaining discrepancies primarily attributed to unavoidable discretization effects.

Although the proposed algorithm relies on periodic multisine data, this requirement mainly supports the frequency-domain estimation of the \ac{bla} and the final optimization stage. The central contribution of this work, the sliding-window estimation of the nonlinear restoring force, is not intrinsically tied to periodic data. As a result, extending the method to arbitrary input-output measurements is relatively straightforward by carrying out the remaining estimation steps directly in the time domain.

\appendix
\section{Computing the sample noise covariance}\label{sec:sample_noise_cov}
Given \(P > 1\), we can compute sample noise covariance matrices that quantify the disturbing noise source \(v(n)\). These matrices are useful as they (i) provide insight into the noise properties, (ii) form the basis for defining weighting matrices in the parameter estimation procedure, and (iii) support model validation.

In the time domain, the sample noise covariance matrix is computed as
\begin{equation}
  \hat{\Sigma}_y^{\text{time}}
  = \frac{1}{NR(P - 1)}
  \sum_{n=0}^{N-1}\sum_{r=0}^{R-1} \sum_{p=0}^{P-1}
    \big(y^{[r,p]}(n) - {y}^{[r]}(n)\big)\big(y^{[r,p]}(n) - {y}^{[r]}(n)\big)^{\mathrm{T}},
  \label{eq:sample_noise_cov_time}
\end{equation}
where \({y}^{[r]}(n) = \frac{1}{P} \sum_{p=0}^{P-1} y^{[r,p]}(n)\) denotes the sample mean over the periods.
In the frequency domain, the sample noise covariance matrix at frequency line \(k\) is computed as
\begin{equation}
  \hat{\Sigma}_{y}^{\text{freq}}(k) = \frac{1}{R (P - 1)} \sum_{r=0}^{R-1} \sum_{p=0}^{P-1}
  \big(Y^{[r,p]}(k) - {Y}^{[r]}(k)\big)\big(Y^{[r,p]}(k) - {Y}^{[r]}(k)\big)^\mathrm{H},
  \label{eq:sample_noise_cov_freq}
\end{equation}
where \({Y}^{[r]}(k) = \frac{1}{P} \sum_{p=0}^{P-1} Y^{[r,p]}(k)\) denotes the sample mean over the periods.
Note that in the time domain, we average over all samples as the noise is assumed stationary, whereas in the frequency domain, each frequency is treated separately since the noise may be frequency-dependent.

\printcredits

\section*{Declaration of Generative AI and AI-assisted technologies in the writing process}
During the preparation of this work, the authors used ChatGPT to assist with language refinement, grammatical corrections, and improvements to clarity and conciseness of the manuscript text. After using this tool, the authors reviewed and edited all content as needed and take full responsibility for the content of the publication.

\bibliographystyle{cas-model2-names}
\bibliography{cas-refs}


\end{document}